\documentclass[10pt,iop]{emulateapj}

\begin{document}

\title{Opening the 21cm EoR Window: Measurements of Foreground Isolation with PAPER}

\author{Jonathan C. Pober\altaffilmark{1}, 
Aaron R. Parsons\altaffilmark{1},
James E. Aguirre\altaffilmark{2}, 
Zaki Ali\altaffilmark{1}, 
Richard F. Bradley\altaffilmark{3,4,5}, 
Chris L. Carilli\altaffilmark{6},
Dave DeBoer\altaffilmark{7},
Matthew Dexter\altaffilmark{7},
Nicole E. Gugliucci\altaffilmark{5}, 
Daniel C. Jacobs\altaffilmark{8},
Patricia J. Klima\altaffilmark{4}, 
Dave MacMahon\altaffilmark{7},
Jason Manley\altaffilmark{9},
David F. Moore\altaffilmark{2}, 
Irina I. Stefan\altaffilmark{10},
William P. Walbrugh\altaffilmark{9}
}
\altaffiltext{1}{Astronomy Dept., U. California, Berkeley, CA}
\altaffiltext{2}{Dept. of Physics and Astronomy, U. Pennsylvania, Philadelphia, PA}
\altaffiltext{3}{Dept. of Electrical and Computer Engineering, U. Virginia, Charlottesville, VA}
\altaffiltext{4}{National Radio Astronomy Obs., Charlottesville, VA}
\altaffiltext{5}{Dept. of Astronomy, U. Virginia, Charlottesville, VA}
\altaffiltext{6}{National Radio Astronomy Obs., Socorro, NM}
\altaffiltext{7}{Radio Astronomy Lab., U. California, Berkeley, CA}
\altaffiltext{8}{School of Earth and Space Exploration, Arizona State U., Tempe, AZ}
\altaffiltext{9}{Square Kilometer Array, South Africa Project, Cape Town, South Africa}
\altaffiltext{10}{Cavendish Lab., Cambridge, UK}

\begin{abstract}
We present new observations with the Precision Array for Probing the Epoch of Reionization
(PAPER) with the aim of measuring the properties of foreground emission for 21cm 
Epoch of Reionization experiments at 150 MHz.  We focus on the footprint of the foregrounds
in cosmological Fourier space to understand which modes of the 21cm power spectrum
will most likely be compromised by foreground emission.  These observations confirm
predictions that foregrounds can be isolated to a ``wedge"-like region of 2D
$(k_{\perp},k_{\parallel})$-space, creating a window for cosmological studies
at higher $k_{\parallel}$ values.  We also find that the emission extends past the nominal
edge of this wedge due to spectral structure in the foregrounds, with this
feature most prominent on the shortest baselines.  Finally, we filter the data to retain
only this ``unsmooth" emission and image specific $k_{\parallel}$ modes of it.  
The resultant images show an excess of power at the lowest modes, 
but no emission can be clearly localized to any one region of the sky.
This image is highly suggestive that the most problematic foregrounds for 21cm EoR
studies will not be easily identifiable bright sources, but rather an aggregate of fainter
emission.
\end{abstract}

\keywords{cosmology: observations --- dark ages, reionization, first stars --- techniques: interferometric}

\section{Introduction}

The highly redshifted 21cm line of neutral hydrogen is widely regarded as one of the most
promising probes of the high redshift universe, with potential to map out volumes
extending from redshift $\sim 1$ through the Epoch of Reionization (EoR) to the dark ages
at redshift 20 and beyond (for reviews of the field, see \citealt{furlanetto_et_al_2006}, 
\citealt{morales_and_wyithe_2011}, and \citealt{pritchard_and_loeb_2012}).
Numerous facilities and experiments targeting the signal from the EoR
are already online or under construction, including 
the LOw Frequency ARray (LOFAR; \citealt{lofar})\footnote{http://www.lofar.org/}, 
the Murchison Widefield Array (MWA; \citealt{mwa})\footnote{http://www.mwatelescope.org/}, 
and the Donald C. Backer Precision Array for Probing the Epoch of Reionization
(PAPER; \citealt{parsons_et_al_2010})\footnote{http://eor.berkeley.edu/}.  
All 21cm cosmology experiments will need to
separate bright galactic and extragalactic foregrounds from the neutral hydrogen signal,
which can be fainter by as much as 5 orders of magnitude or more (see, e.g.,
\citealt{furlanetto_et_al_2006} and \citealt{santos_et_al_2005}).  

Considerable effort has been devoted to developing schemes to remove or isolate foregrounds
from 21cm data (e.g. \citealt{morales_et_al_2006}, \citealt{bowman_et_al_2009}, 
\citealt{liu_et_al_2009}, \citealt{liu_and_tegmark_2011}, \citealt{parsons_et_al_2012b},
\citealt{dillon_et_al_2012}).
Almost all of these approaches rely on the spectral smoothness of foreground emission
relative to the 21cm signal, which will contain significant structure versus frequency.
The purpose of this letter is to use the delay transform technique presented in
(\citealt{parsons_et_al_2012b}; hereafter, P12b) on
observations from PAPER to test the behavior of actual foreground emission.
We wish to understand the footprint of foregrounds in $k$-space
to determine which modes of the 21cm power spectrum will be most accessible to observation.
The structure of this letter is as
follows: in \S\ref{sec:data}, we describe the data used in these observations.
In \S\ref{sec:analysis}, we review the delay spectrum technique presented in
P12b, and then describe the steps used in applying this approach to actual
observations.  We present our results in \S\ref{sec:results} and conclude in 
\S\ref{sec:conclusions}.

\section{The Data}
\label{sec:data}

\begin{figure}\centering
\includegraphics[height=3in]{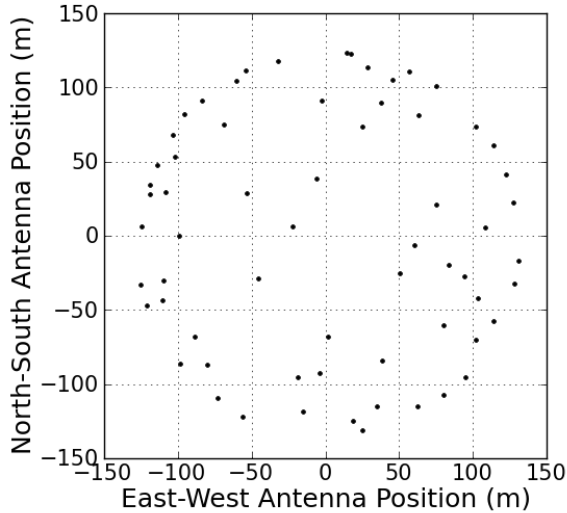}
\caption{
The configuration of the 64 PAPER dipoles used in this analysis.  The zero-point
is the center of the array.  The y-axis is North/South, the x-axis is East/West, and distances 
are in meters.
}
\label{fig:antenna_layout}
\end{figure}

We use 4 hours of data collected in 10 second integrations between JD 2455747.48 and 2455747.64 
($4-5$ July 2011), using 
a 64 element PAPER array located on the SKA site in the Karoo region of South Africa.
This data set comes from the same observing campaign described by \citet{stefan_et_al_2012},
although this specific 4-hour window falls outside of the observations analyzed therein.
The dipole antennas are arranged in a
``minimum redundancy" configuration optimized for imaging analysis
\citep{parsons_et_al_2012a}, which is shown
in Figure \ref{fig:antenna_layout}.  
This configuration has a maximum baseline
length of $\sim 300$m, corresponding to an image plane resolution of $0.4^\circ$
at 150 MHz.
The PAPER correlator has a 100 MHz instantaneous bandwidth 
from $100-200$~MHz divided into 2048 frequency channels.
We correlate only one linear polarization on each dipole, and
discard all data from antennas 40 and 55 which were cross-polarized. 

\begin{figure*}\centering
\includegraphics[width=7in,trim=4cm 0cm 4cm 0cm]{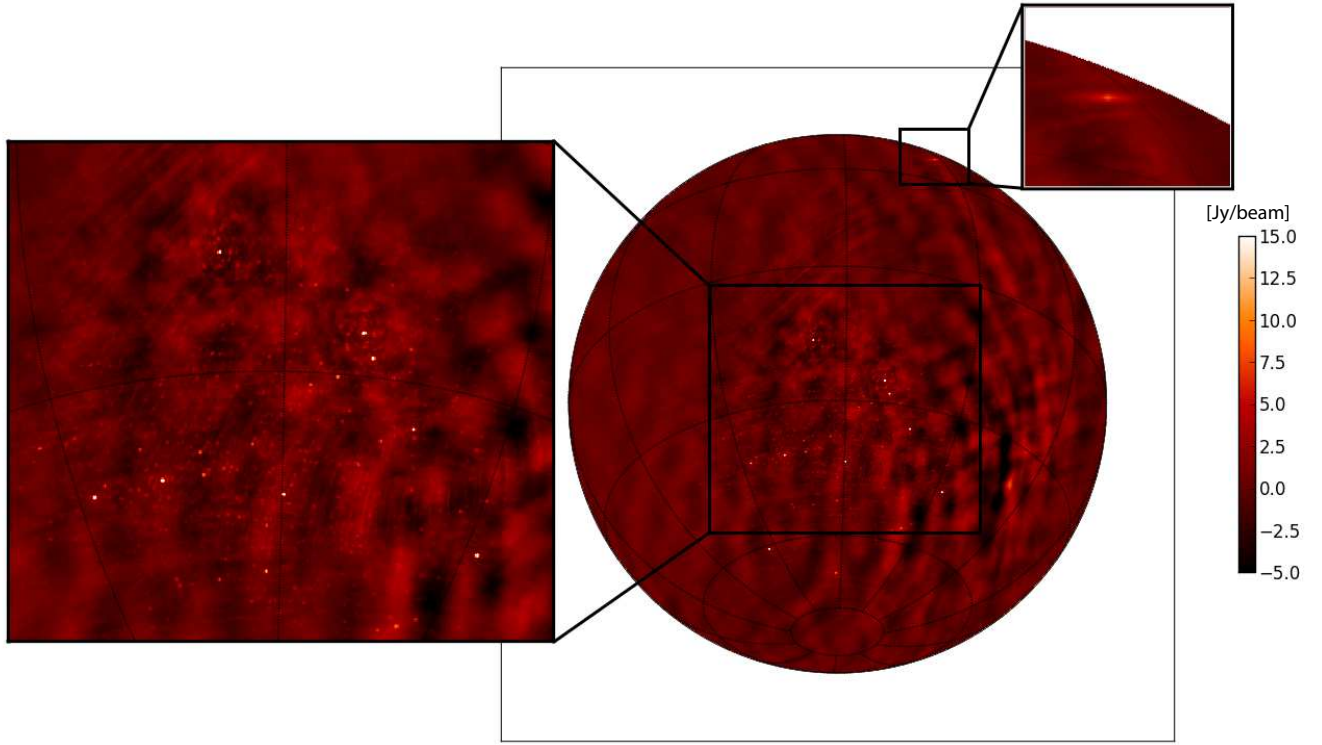}
\caption{
A dirty image of the data used in this analysis, centered on RA 
21h52m and declination -30$^\circ$43', the transiting zenith halfway through the observation.
Prominent sidelobes from the Galaxy are seen in the right-hand side of the map.  Close-ups
show Cygnus A at 19h59m and +40$^\circ$44' and the point source population of a potential
EoR cold-patch. The color-scale is linear in Jy, with only the brightest point sources saturating the
range.}
\label{fig:skymap}
\end{figure*}

An image of the field transiting during this time period is shown in Figure \ref{fig:skymap}.
This image spans $140 - 165$ MHz; 100 sub-bands of 0.25 MHz
were individually imaged and summed to make the map shown.
No CLEANing or spectral slope correction was performed.
The observation is centered on a low-foreground ``cold patch," a potential field for an
EoR science observation.  The Galactic plane, which is just setting at the end
of the 4-hour observation, creates prominent sidelobes over the entire map.

Complex antenna based gains are derived
using fringe-fitting to Centaurus A, Pictor A, 
and Fornax A;
an overall gain scale is derived from the \citet{helmboldt_et_al_2008} source J2214-170.
We perform a small gain linearization
correction to mitigate data quantization effects in the correlator (described in 
\citealt{parsons_et_al_2010}) and a correction for temperature dependent gain drifts
(described in \citealt{pober_et_al_2012}).  We also perform
radio-frequency interference (RFI) excision, manually flagging frequency channels
of known transmitters, and flagging any points $6\sigma$ above the mean after
differencing adjacent channels along the frequency axis.

\section{Analysis Techniques}
\label{sec:analysis}

At the core of our analysis is the delay spectrum technique presented in
P12b.  Since cosmological redshifting maps the observed 21cm
line frequency into a distance measurement, the Fourier transform of the
frequency axis are the $k_{\parallel}$ line-of-sight modes of the 21cm power spectrum.  
However, this relation is clearly not true for foregrounds, where the frequency axis
simply corresponds to the spectra of the sources.
The delay transform first presented in \citet{parsons_and_backer_2009} provides
a framework for mapping foreground emission into a cosmological $k$-space.
The frequency Fourier transform of a single baseline's visibility spectrum (the ``delay transform")
maps celestial emission to ``delay space," 
where sources appear as Dirac delta functions, located at the
delay in arrival time between the two elements of that baseline.  
These delays must be limited to values below the physical light travel time between
the two antennas (the ``horizon limit"), and so
all emission from the sky maps to a region in the center of delay space determined
only by the baseline length.
However, any spectral structure in the visibilities acts as a convolving kernel
in delay space.  Foreground emission
is spectrally smooth, translating into a narrow convolving kernel; 21cm emission has
large amounts of spectral structure, and therefore its kernel scatters power from within
the horizon limit to larger delays.
P12b also showed that delay has a near one-to-one
mapping to $k_{\parallel}$, meaning that those delay modes free from
contaminating foreground emission are effective probes of the 21cm power spectrum.
The baseline-length dependence creates better isolation on the shortest baselines, giving
rise to the ``wedge" structure seen in 
\citet{datta_et_al_2010}, \citet{vedantham_et_al_2012}, \citet{morales_et_al_2012}, 
\citet{trott_et_al_2012}, and P12b; we refer the reader to these works for more detailed
and alternative derivations of the ``wedge."

\subsection{Delay Space CLEAN}
\label{sec:clean}

Some of the practical aspects of implementing the delay-spectrum approach in actual data
were described in \S3 of P12b.  Of particular importance is the
implementation of the frequency Fourier transform using 
a window function and the 1D-CLEAN algorithm first presented in 
\citet{parsons_and_backer_2009} to reduce the effects of RFI flags and band-edge effects.
Even if foregrounds are spectrally smooth and easily localized in 
delay/$k_{\parallel}$-space, such sharp edges in frequency space 
will introduce significant covariance in delay space, 
resulting in the scattering of foreground emission into otherwise uncontaminated
regions of $k$-space. The 1D-CLEAN algorithm
treats RFI flags as a sampling function in frequency space,
and ``fills in" these gaps by iteratively fitting the brightest Fourier components 
in the delay domain.
As the sampling function is known exactly, this algorithm has proven extremely
effective at removing covariance between delay modes.  
The end result is a model of our data
which is free of RFI flagging gaps.  We then 
form power spectra both of this model and the residuals between it and the raw data,
before re-combining them.  By separating the two components before the
Fourier transform, we reduce the amount of power that can scatter off RFI gaps,
minimizing bright sidelobes which would otherwise contaminate the EoR window
in the power spectrum.

We force our foreground model to be smooth-spectrum in
frequency by only allowing delay-space components which fall inside 
a baseline-dependent area (a ``CLEAN box''), 
50~ns beyond the physical maximum horizon delay on that baseline.
This extra 50~ns allows the algorithm to model foreground emission pushed beyond the
horizon limit.  
For this work, 50~ns appears to encompass enough foreground emission
that the sidelobes of any remaining flux scattering off RFI gaps 
in the residuals are below
the noise.  Since our results do detect
additional emission beyond our chosen cut-off, its exact value 
may need to be revisited in future analyses with more sensitivity.

\subsection{Power Spectra}
\label{sec:pspec}
Once the data have been CLEANed, we 
form power spectra on a visibility-by-visibility basis.  Our power spectrum estimates
follow from equation (12) of \citet{parsons_et_al_2012a}:
\begin{equation}
\label{eq:pk}
\widehat P(k) \approx \tilde V_{21}^2 \left(\frac{\lambda^2}{2k_{\rm B}}\right)^2 \frac{X^2Y}{\Omega B},
\end{equation}
where $\lambda$ is the observing frequency, $k_{\rm B}$ is Boltzmann's constant, $\Omega$ is the 
solid angle of the primary beam,\footnote{Derivations of equation \ref{eq:pk} in
\citet{morales_2005}, \citet{mcquinn_et_al_2006}, and \citet{parsons_et_al_2012a}, 
relate $\Omega B$ to an effective cosmological
volume, using a top-hat primary beam or effective area as a pedagogical tool
to simplify the result.  More generally,
however, the effective $\Omega$ in equation \ref{eq:pk} is actually
${\int A^2(\theta,\phi)\ \rm{d}\theta\ \rm{d}\phi}$, where $A(\theta,\phi)$ is the 
power response of the primary beam.
A derivation of this effect will be presented in a subsequent full length publication
by Parsons et al.
For PAPER, this beam is a factor of $\sim2$ smaller than ${\int A(\theta,\phi)\ \rm{d}\theta\ \rm{d}\phi}$.} 
$B$ is the observing bandwidth, $X$ and $Y$ are cosmological
scalars which convert observed angles and frequencies into $h{\rm Mpc}^{-1}$, and $\tilde V$ is
a delay transformed visibility.
We avoid introducing a noise-bias by forming
our estimator $\tilde V_{21}^2$ from adjacent time samples on a baseline.
The 10-second interval between integrations is short enough that both measurements can be
considered redundant samples of the same $k$-modes.

The isotropy of the universe allows us to then combine all power spectrum estimates 
$\tilde V_{21}^2$ in annuli
of equal $k_{\perp}$ to form a 2D power spectrum in the $(k_{\perp},k_{\parallel})$-plane.
We note that the method used here does not take advantage of any coherent integration within 
a $uv$-pixel.  Since foreground emission dominates the observed signal, the loss
of sensitivity is not important, and we ignore this effect for
computational efficiency.  For EoR science runs, however, 
PAPER explicitly focuses on maximizing the sensitivity boost from coherent integration,
using ``maximum redundancy" configurations designed to sample select
$uv$-pixels for long periods of time.

\section{Results}
\label{sec:results}

To make power spectra of the 4-hour dataset described above, we first run our CLEAN algorithm
over the full 100~MHz band.  Using the whole band gives the best resolution in delay space,
and since foregrounds are nearly coherent over the whole band, the additional information
gives CLEAN the most signal-to-noise to build its model.  For the cosmological delay
transform to make power spectra, we use only a 25 MHz band from 140~to~165~MHz.  This smaller
band still exceeds the $\sim8$~MHz band over which the $z\sim8$ universe can be treated as
coeval \citep{furlanetto_et_al_2006}.  
Since the main purpose of this work is to understand the $k$-space behavior of foregrounds,
we ignore this effect, as the additional bandwidth gives us better $k$-space resolution.

\begin{figure*}\centering
\includegraphics[width=7in,trim=0cm 0cm 3cm 0cm]{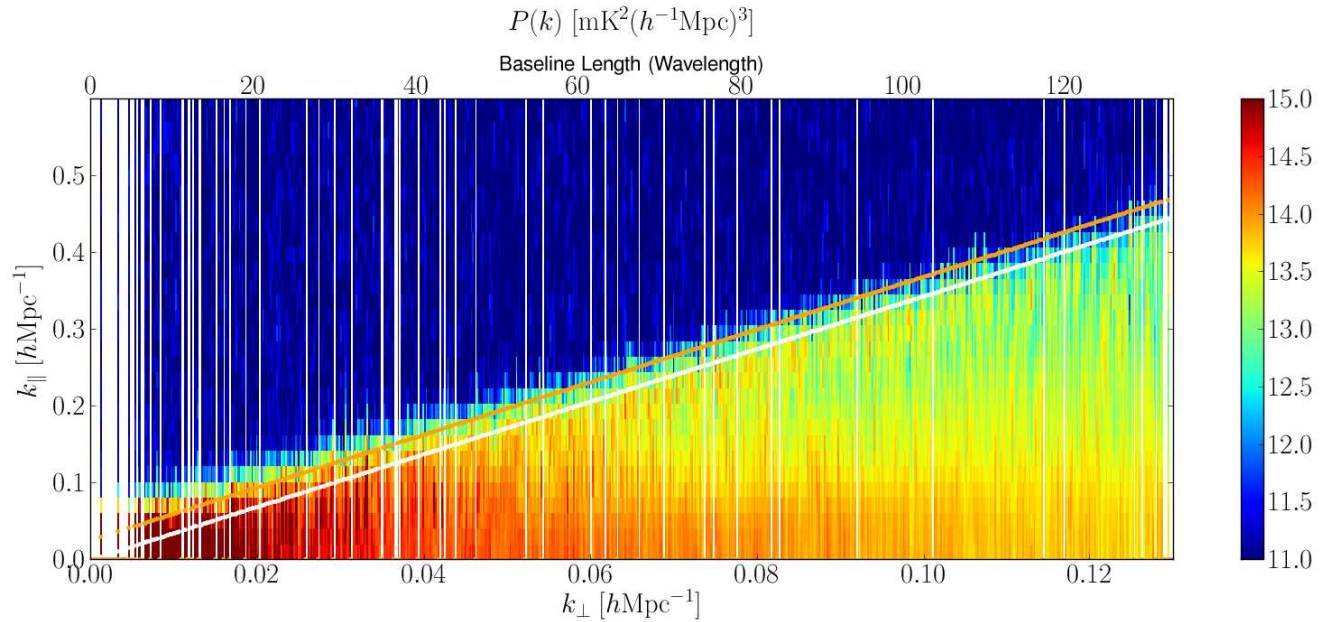}
\caption{
A two-dimensional power spectrum of the 4 hours of data analyzed.  The wedge-like nature of the
foreground emission is clear.
The white line marks the horizon limit and the orange line is 50~ns beyond.  
The colorscale is logarithmic and the units are ${\rm mK}^2~(h^{-1}\rm{Mpc})^3$
The binning is described in the text.}

\label{fig:wedge}
\end{figure*}

Forming individual power spectra from each baseline of the array and binning in
$|k_{\perp}|$ yields the 2-dimensional $P(k_{\perp},k_{\parallel})$ shown in Figure \ref{fig:wedge}.
The $k_{\perp}$-axis is binned with a resolution of $1.87\times10^{-4} h{\rm Mpc}^{-1}$; 
gaps are visible where there are no baselines of that length.
We do no binning in $k_{\parallel}$; this resolution is set by the 25~MHz bandwidth used in
the analysis.\footnote{Because of the Blackman-Harris window function used in CLEAN,
only every other sample plotted in $k_{\parallel}$ is statistically independent.}
The most prominent feature is the ``wedge"-like shape of the foreground emission 
as predicted.  As argued
in P12b, this ``wedge" footprint in $k$-space is not a result of
imperfect calibration in foreground removal (since we attempt no 
foreground removal in this work), but a property of the emission itself as measured
by an interferometer.

The white diagonal line in Figure \ref{fig:wedge}
corresponds to the horizon-limit in $k_{\parallel}$ for a baseline of corresponding length
$k_{\perp}$; the orange line is 50~ns beyond the horizon, inside of which
we allowed Fourier components in the deconvolution described in \S\ref{sec:clean}.
As predicted, emission extends beyond the horizon limit
due the intrinsic spectral structure of
the foreground emission. 

We draw attention to the fact that the supra-horizon emission does not have a constant
width in $k_{\parallel}$ as a function of $k_{\perp}$.  Rather, more emission extends beyond
the horizon on the smallest $k_{\perp}$-values (i.e. the shortest baselines).  
We expect this behavior to result from two different effects.
First, the shortest baselines will resolve out less of the diffuse 
Galactic synchrotron, so that the emission will be brighter.  Therefore,
we can see its sidelobes extend further in $k_{\parallel}$ before they fall 
below the noise level.
The second effect is somewhat more subtle and can be best illustrated with an example.
Consider two east-west baselines of length 10 and 100$\lambda$ at 150 MHz, 
which correspond to light travel times (i.e. horizon limits) of 66.7 and 667~ns, respectively.  
A point source $20^{\circ}$ above the eastern horizon
corresponds to geometric delays of 62.6 and 626~ns on these baselines.
If the source spectrum has a given amount of ``unsmoothness'' and
creates a delay space kernel of width 10~ns, then this kernel, centered at the geometric
delay of the source, will lead to emission beyond
the horizon on the $10\lambda$ baseline, but not on the $100\lambda$ baseline.
This example illustrates how the same sources of emission will naturally 
lead to more corruption of supra-horizon delays on shorter baselines than on longer ones.

Finally, we draw attention to the ``edge brightening" 
of the foreground emission in the wedge on
the longest baselines as one moves near the horizon limit. This feature can be attributed to
the Galactic plane, and moves as expected when the data are viewed as a function of time.

\begin{figure*}\centering
\includegraphics[width=3.5in,trim=0cm 0cm 0cm 0cm]{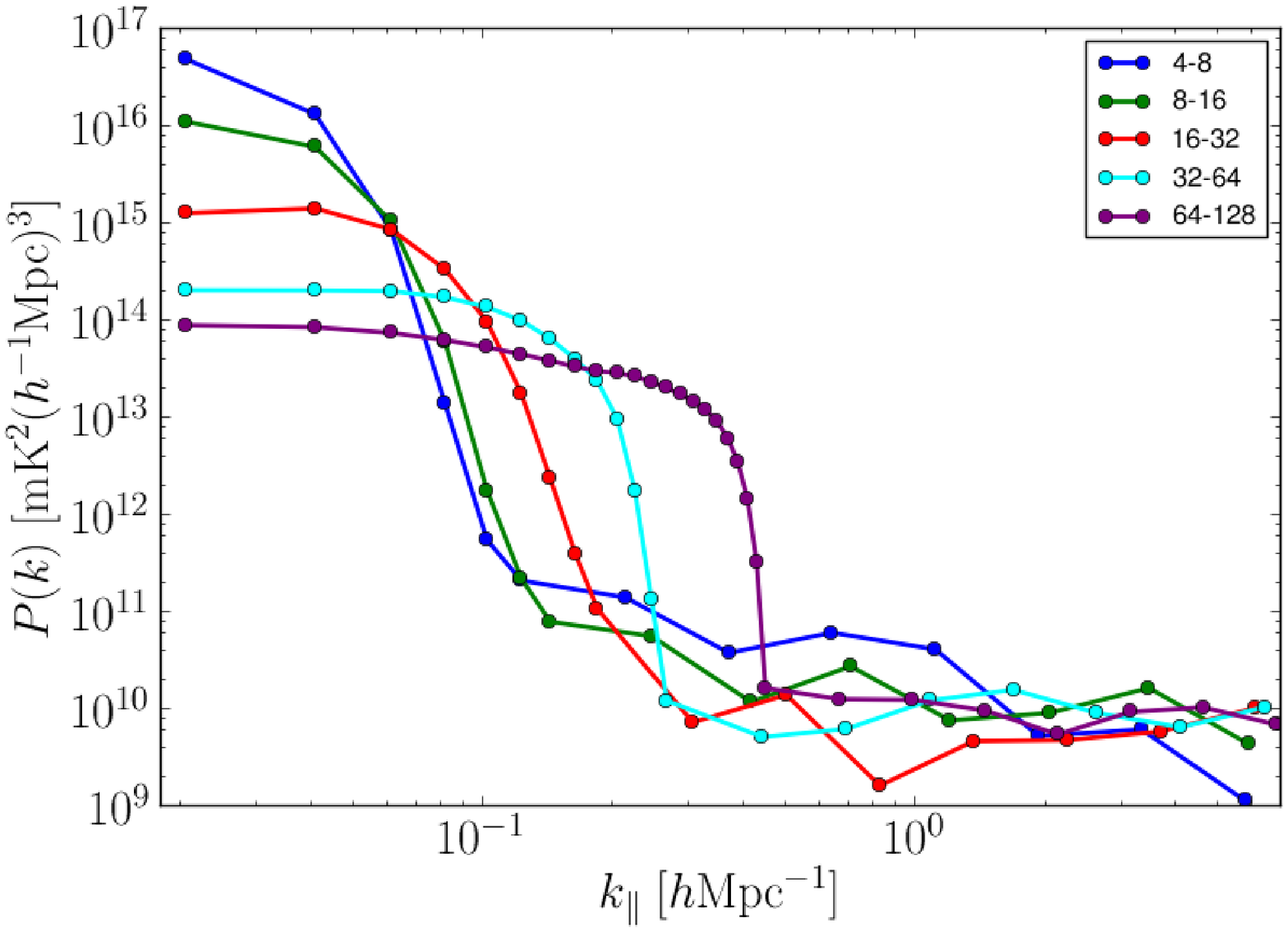}\includegraphics[width=3.5in,trim=0cm 0cm 0cm 0cm]{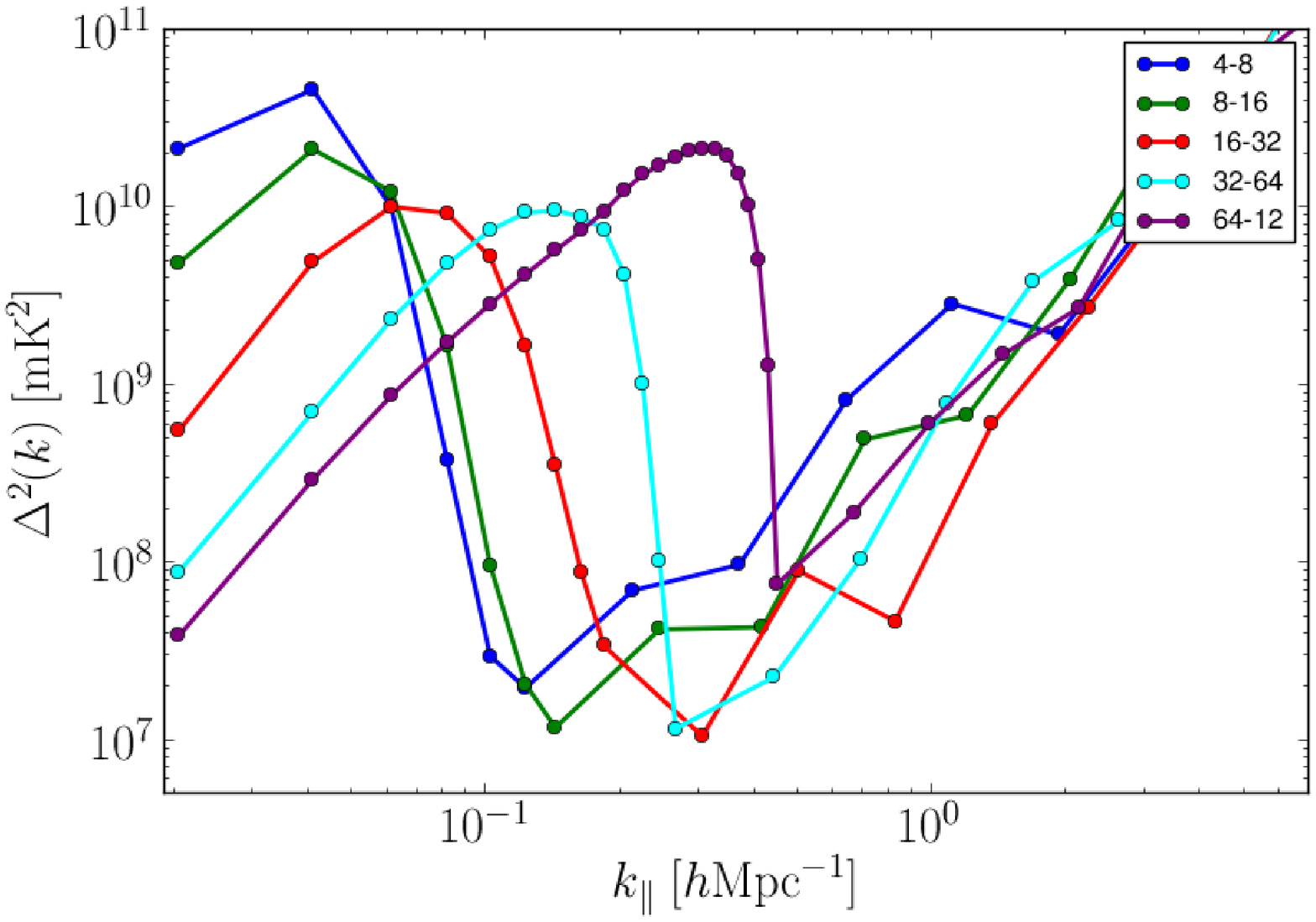}
\caption{
{\emph Left}: A 1-dimensional power spectrum versus $k_{\parallel}$ for bins of several baseline lengths.  
To preserve the steep roll-off of the foreground
emission, the data are plotted with their natural resolution at lower $k_{\parallel}$ values,
and logarithmically binned at higher values. {\emph Right}: The same as the left, but 
the dimensionless power spectrum $\Delta^2(k) = \frac{k^3}{2\pi^2}P(k)$.
}
\label{fig:pspec1d}
\end{figure*}

To highlight the steepness of the foreground roll-off, we plot 
1-dimensional $k_\parallel$ power spectra for bins of several baseline lengths 
in Figure \ref{fig:pspec1d}.  
We see that the foreground emission can fall by as 
much as three to four orders of magnitude in a factor of 2 change in $k_{\parallel}$.  It
is difficult to explicitly compare this result to the predictions of P12b,
due to the different resolutions and binning used.  The placement of bin edges can significantly
complicate comparison when the fall-off is so steep, as a slight shift in the bin can result
in a large change in the average value within.  For similar reasons, it is difficult to say
exactly where the emission falls below the noise.  
Given these caveats, there is nothing in these
data to contradict P12b.  It is clear, however,
that sensitivities will have to increase significantly before anything
can be said about the behavior of foreground emission at the tens of mK$^2$ level where the
expected EoR signal lies. 

\begin{figure*}\centering
\includegraphics[width=7in]{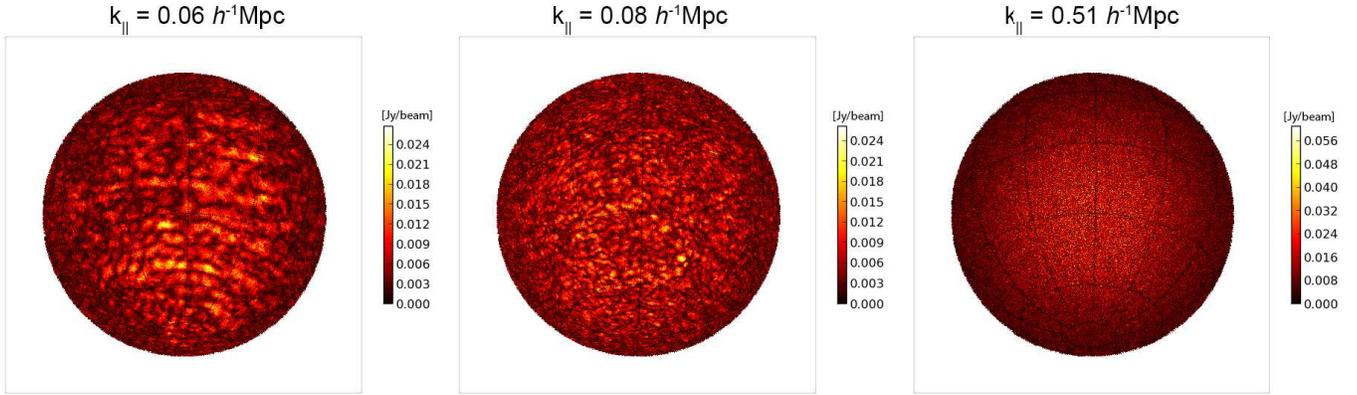}
\caption{
Images of the magnitude of select $k_{\parallel}$ modes 
in the data, after a high-pass delay space filter has removed all emission
interior to 50~ns beyond the horizon limit 
(i.e. emission below the orange line in Figure \ref{fig:wedge} has been removed).  
The color-scale is linear in Jy, although the flux scale 
on the lowest $k_{\parallel}$-modes is compromised
due to the filter.
While emission is clearly present in the low $k_{\parallel}$ modes,
it cannot be identified with the bright sources in Figure \ref{fig:skymap},
nor are the features obviously correlated with Galactic structure.}
\label{fig:fmodes}
\end{figure*}

To both demonstrate the effectiveness of the delay-space foreground isolation
and to further investigate the nature of the supra-horizon emission, 
we high-pass filter the data 
in delay-space, selecting only delay modes more than 50~ns beyond the horizon limit,
i.e., we select the emission lying beyond the orange line
in Figure \ref{fig:wedge}.
We then image this data from 140 to 165 MHz in one-hundred 0.25 MHz bins to form a data-cube
versus frequency.  Finally, we Fourier transform our data-cube versus frequency to create maps of
individual $k_{\parallel}$ modes.
Three of the resultant maps for $k_{\parallel} = 0.06, 0.08,$ and $0.51~h\rm{Mpc}^{-1}$ are shown in
Figure \ref{fig:fmodes}.

The flux scale in these images has been reduced by 2 to 3 orders of magnitude from
Figure \ref{fig:skymap}, demonstrating the effectiveness of delay-space filtering.  Interpreting the flux scale in the lowest $k_{\parallel}$ modes 
is complicated, since most of
baselines at these modes have been filtered off, whereas,
the $k_{\parallel} = 0.51 h\rm{Mpc}^{-1}$ mode is effectively a noise map.
It can be seen in Figure \ref{fig:wedge} that this mode lies above the wedge, and therefore nothing
has been filtered from it.  The RMS in this map (calculated from the complex data) is
14 mJy.  
We can estimate the expected noise level using:
\begin{equation}
\Delta\sigma = \frac{2k_B\Omega}{\lambda^2}\frac{T_{\rm sys}}{\sqrt{N(N-1)Bt}},
\end{equation}
where $k_B$ is Boltzmann's constant, $\Omega$ is the solid angle of the primary beam,
$\lambda$ is the observing wavelength, $T_{\rm sys}$ is the system temperature, $N$ is the
number of antennas in the array, $B$ is the bandwidth, and $t$ is the observing time
\citep{thompson_et_al_2007}.
For this observation $\Omega$ = 0.75~sr, $t = 4$ hours, $\lambda = 1.96$~m, 
$B = 25$~MHz, and $T_{\rm sys}$ = 1000~K.
This somewhat high value for $T_{\rm sys}$ is reasonable given the
Galactic emission in these observations;
the values of
$T_{\rm sys}$ for PAPER will be presented in a forthcoming publication by Parsons et al.
Using these values gives an expected RMS of 14~mJy, in accord with
our measurement.

The $k_{\parallel} = 0.06$ and $0.08~h\rm{Mpc}^{-1}$ maps are clearly not noise dominated.
Accounting for the noise that was removed by filter, these data have rough
``effective" RMS values of 35~mJy, well
in excess of our noise estimate.  Given the presence of emission in these modes above the 
orange line in Figure 
\ref{fig:wedge}, it is not surprising that we see excess power.  
What is surprising is that none of this emission can be easily
associated with the brightest sources or structures visible in Figure \ref{fig:skymap}.  
Rather, it appears that the bulk of the emission contaminating the EoR window comes 
from an aggregate of fainter emission.
One might be concerned that emission can leak from lower $k_{\parallel}$ bins
into the modes shown, but as argued in \S\ref{sec:clean}, our deconvolution
algorithm is extremely effective at eliminating this covariance.
However, we caution the reader against interpreting these maps as images of true coherent structures
on the sky.  The fact that the spatial pattern of emission changes significantly from 
$k_{\parallel} = 0.06$ to $0.08~h\rm{Mpc}^{-1}$ suggests the emission is a diffuse background and
our images are limited by sidelobes.

We note one additional interesting feature in these maps.  The noise-dominated $k_{\parallel} = 0.51~h\rm{Mpc}^{-1}$ gives a good impression of the PAPER primary beam shape.  
In the 0.06 and $0.08~h\rm{Mpc}^{-1}$ maps, however, bright emission extends well beyond the half-power point of
the beam (roughly $45^{\circ}$ FWHM).  Under the lens of the delay
transform, one suspects that the emission closest to the horizon can most easily
create supra-horizon emission.  These images show that while emission from well outside
the beam FWHM contributes significantly to the supra-horizon emission, the PAPER primary beam
roll-off is enough to keep this pattern from extending all the way out the edge of the image.  

\section{Conclusions}
\label{sec:conclusions}

We have presented new observations from PAPER measuring the properties of 
foreground emission
in cosmological Fourier space.  These observations have confirmed general predictions
presented in, e.g., \citet{datta_et_al_2010},
\citet{morales_et_al_2012} and P12b: that foreground emission
occupies a ``wedge" in the 2D $(k_{\perp},k_{\parallel})$ plane, leaving a window at
higher $k_{\parallel}$ values for 21cm EoR studies.  We have also confirmed
that shorter baselines
yield a larger window onto the cosmological signal.  However, this the
window does not grow perfectly linearly with decreasing baseline length.  
Therefore, while shorter
baselines do make the best probes of the EoR signal, there will be diminishing returns at the
shortest baselines.
We have also presented an images of several $k_{\parallel}$ modes of
``unsmooth" emission extending past
the nominal edge of the wedge.
These images are unable to localize any of the emission to known sources on the sky,
suggesting that the most problematic foregrounds for EoR observations are a diffuse
background.

\acknowledgments{We would like to thank our reviewer for their thoughtful and helpful comments.
We thank SKA-SA for their efforts
in ensuring the smooth running of PAPER. PAPER is
supported through the NSF-AST program (awards 0804508, 1129258, and 1125558), 
the Mt. Cuba Astronomical Association, and by
significant efforts by staff at NRAO.
}

\end{document}